\documentclass{webofc}
\usepackage[varg]{txfonts}   
\newcommand{\hm}[1]{#1\nobreak\discretionary{}{\hbox{\ensuremath{#1}}}{}}
\newcommand{\myfrac}[2]{{\ifmmode{}^{#1}\!/_{\!#2}\else${}^{#1}\!/_{\!#2}$\fi}}

\newcommand{\beq}{\begin{equation}}
\newcommand{\eeq}{\end{equation}}
\newcommand{\beqar}{\begin{eqnarray}}
\newcommand{\eeqar}{\end{eqnarray}}

\def\dalam{\hbox
{\vrule\vbox{\hrule\hbox to 1ex{ \hfill}\kern 1 ex\hrule}\vrule}}

\def\1/2{\hbox{$ {1 \over 2}$ }}

\def\h{\hbar}
\def\i/h{{i \over \h}}

\def\ch{\cosh}
\def\sh{\sinh}

\def\a{\alpha}

 \def\D{\Delta}

\def\<{\langle}
\def\>{\rangle}

\def\({\left(}
\def\[{\left[}
\def\){\right)}
\def\]{\right]}

\begin{document}
\title{Estimating the radiative part of QED effects in systems with supercritical charge}
%
%

\author{\firstname{Artem} \lastname{Roenko}\inst{1,2}\fnsep\thanks{\email{roenko@physics.msu.ru}} \and
        \firstname{Konstantin} \lastname{Sveshnikov}\inst{1}\fnsep\thanks{\email{costa@bog.msu.ru}} 
}

\institute{Faculty of Physics, Lomonosov Moscow State University, Leninskie Gory, Moscow 119991, Russia
\and
           Bogoliubov Laboratory of Theoretical Physics, Joint Institute for Nuclear Research, Dubna 141980, Russia
          }

\abstract{%
The effective interaction of the electron magnetic moment anomaly with the Coulomb field of superheavy nuclei is investigated by taking into account its dynamical screening at small distances. The shift of the electronic levels, caused by this interaction, is considered for H-like atoms and for compact nuclear quasi-molecules, non-perturbatively both in $Z\alpha$ and (partially) in $\alpha/\pi$. It is shown that the levels shift reveals a non-monotonic behavior in the region $Z\alpha>1$ and near the threshold of the lower continuum decreases both with the increasing the charge and with enlarging the size of the system of Coulomb sources. The last result is generalized to the total self-energy contribution to the levels shift and so to the possible behavior of radiative QED effects with virtual photon exchange near the lower continuum in the supercritical region.
}
\maketitle
%
\section{Introduction}
\label{sec:intro}
For the supercritical region, $Z>Z_{cr}\simeq 170$, QED predicts the non-perturbative vacuum reconstruction due to diving of the discrete electronic levels into the lower continuum, which should be followed by the vacuum positron emission~\cite[and refs. therein]{Greiner1985a, Popov2001}. In addition, the recent essentially non-perturbative computations show that the vacuum polarization energy demonstrates an substantially non-linear behavior in the supercritical region,  gradually different from the predictions of the perturbation theory; moreover, under certain conditions, the vacuum polarizaton energy can compete with the classical electrostatic energy of the Coulomb sources~\cite{Davydov2017, Sveshnikov2017, Davydov2018a}. Hence, an actual problem is to study  the possibility of compensating these effects caused by fermionic loops by virtual photon exchange.

Since the completely non-perturbative analysis of the radiative QED effects lyes beyond the existing methods of  computation, the investigation of those separate components of the self-energy shift, which allow for a detailed analysis, turns out to be of a special interest. One of these effects is the interaction $\Delta U_{AMM}$ of the electron's magnetic moment anomaly (AMM), dynamically screened at small distances, with the Coulomb field of nuclei, which is described by a local operator and so preserves all the requirements for the Furry picture.


The behaviour of the electronic levels near the threshold of the lower continuum, caused by $\Delta U_{AMM}$, is considered for H-like atom and compact nuclear quasi-molecules, which simulate the low-energy heavy-ions collisions. The dependence of the levels shifts $\Delta E_{AMM}$ on the charge $Z$ of the nuclei and on the internuclear distance $d$ is investigated. It is shown, that for the levels located in the vicinity of the lower continuum threshold the magnitude of $\Delta E_{AMM}$ decreases both with growing $Z$ and $d$.



\section{The Dirac equation with $\Delta U_{AMM}$}
\label{sec:1}
\subsection{The effective interaction caused by the electron's magnetic anomaly}
\label{sec:1a}
The effective interaction of the AMM with an external electromagnetic field by taking into account the dependence of the electronic form factor $F_2(q^2)$ on the momentum transfer $q^2$ from the very beginning is given by the following expression~\cite{Lautrup1976, Roenko2017}
\begin{equation}\label{eq:f3}
\Delta U_{AMM}(\vec{r}\,)= \frac{e}{ 2m}\, \sigma^{\mu\nu}\partial_\mu \mathcal{A}^{(cl)}_\nu(\vec{r}\,) ,
\end{equation}
where $e$ is an electron charge, $m$ is an electron mass,
\begin{equation}\label{eq:f4}
\mathcal{A}^{(cl)}_\mu(\vec{r}\,) = \frac{1}{(2\pi)^3} \int \! d\vec{q} \ e^{i \vec{q}\,\vec{r}}\, \tilde{A}_\mu^{(cl)}(\vec{q}\,) F_2(-\vec{q}\,^2) \, ,
\end{equation}
while $\tilde{A}_\mu^{(cl)}(\vec{q}\,)$ is the Fourier-transform of the classical external field $A_\mu^{(cl)}(\vec{r}\,)$. In the static limit $F_2(q^2)\simeq F_2(0)$ from the Eq.~\eqref{eq:f3} one obtains the well-known Dirac-Pauli operator.
The effective potential~\eqref{eq:f3} of the interaction with one or more spherically-symmetric extended nuclei with charge $Z_i$ and coordinate $\vec{r}_i$ could be rewritten as the following commutator~\cite{Roenko2017, Roenko2017c}
\begin{equation}\label{eq:06}
\D U_{AMM}(\vec{r}\,)= -\lambda \,\left[ \vec{\gamma} \cdot \vec{p}\,,\,V(\vec{r}\,)\right] \, ,
\end{equation}
where $\lambda=\alpha^2/4\pi m$, $\alpha=e^2/4\pi$, $V(\vec{r}\,)=\sum_i Z_i \, {c\left(|\vec{r}-\vec{r_i}|\right)}/{|\vec{r}-\vec{r_i}|}$, while the function $c(r)$ for the nucleus in the form of the uniformly charged ball with radius $R$ is given by~\cite{Roenko2017}
\begin{equation}\label{eq:s16}
\begin{split}
c(r)&=1- \int\limits_{4m^2}^\infty \! \frac{dQ^2}{Q^2}\, \frac{3QR\ch QR - 3\sh QR}{R^3 Q^3} \, e^{- Q r}\, \frac{1}{\pi}\, \frac{\text{Im}\, F_2(Q^2)}{F_2(0)},\qquad r>R\, ,\\
c(r)&=\frac{(3 R^2 - r^2)}{2 R^3}\,r - \frac{r}{2 m^2 R^3} +{} \\
{}&+ \int\limits_{4m^2}^\infty \! \frac{dQ^2}{Q^2}\, \frac{3(QR+1)}{R^3 Q^3}\,\sinh Qr \, e^{- Q R}\, \frac{1}{\pi}\, \frac{\text{Im}\, F_2(Q^2)}{F_2(0)} , \qquad\qquad \qquad \,\, r<R\, .
\end{split}
\end{equation}
where in the one-loop approximation $ \dfrac{1}{\pi}\, \text{Im}\, F_2(Q^2) = 2 F_2(0)\, \dfrac{m^2}{Q^2} \, \dfrac{1}{\sqrt{1-4m^2/Q^2}}$.

The expression $\Delta g_{free} \, c(r)$, where $\Delta g_{free}=2 F_2(0)\simeq \alpha/\pi$ is an AMM of the free electron, could be interpreted as the dependence of the electronic AMM on the distance from the nucleus center~\cite{Lautrup1976}, and the examination of expression~\eqref{eq:s16} shows that $c(r) \to 1$ in the region $r \gtrsim 1/m$, while for $r\to 0$ it tends to zero. Thus, accounting for the 
form factor dependence on the momentum transfer  the Eq.~\eqref{eq:f4} ensures suppression of the AMM-potential at small distances, principally important for superstrong fields~\cite{Lautrup1976,Roenko2017a, Roenko2017b}.

%
%
\subsection{The one-center and two-center Dirac equation}
\label{sec:1b}
The Dirac equation with an additional 
interaction $\Delta U_{AMM}$ takes the form ($\hbar=c=m_e=1$)
\begin{equation} \label{eq:1}
\left( \vec{\alpha}\vec{p} + \beta + W (\vec{r}\,) + \Delta U_{AMM}(\vec{r}\,) \right) \psi = \epsilon \psi \, ,
\end{equation} 
where $W(\vec{r}\,)$ is the Coulomb interaction of the electron with the nuclei. For our purposes it is convenient to write $W(\vec r\,)$  in the form $W(\vec{r}\,) = -\alpha \, U(\vec{r}\,)$, where 
$U(\vec{r}\,) = \int d\vec{r}\,'\ {\rho(\vec{r}\,)}/{|\vec{r}-\vec{r}\,'|}$,
while $\rho(\vec{r}\,)\hm = \sum_i \rho_0(\vec{r}-\vec{r}_i\,)$, and in the considered nuclear model   
$ \rho_0(\vec{r}\,)=\Theta(R-r)\, {3Z_i}/{4\pi R^3}$.

From~the~Eq.~\eqref{eq:1}~for~the~upper~$i\varphi$~and~lower~$\chi$~components~of~the~Dirac~bispinor $\psi$ there follows
\begin{equation}\label{eq:6}
\begin{split}
i \left( \vec{\sigma}\vec{p} + \lambda \left[ \vec{\sigma} \vec{p} \,, V (\vec{r}\,) \right] \right) \varphi & = \left( \epsilon + 1 + \alpha  U(\vec{r}\,) \right) \chi \, ,\\
i \left( \vec{\sigma}\vec{p} - \lambda \left[ \vec{\sigma} \vec{p} \,, V (\vec{r}\,) \right] \right) \chi & = - \left( \epsilon - 1 + \alpha  U(\vec{r}\,) \right) \varphi  \, .
\end{split}
\end{equation}

Let us consider one-center and two-center configurations of the Coulomb sources corresponding to the H-like atom and to the simplest nuclear quasi-molecule, which consists of two identical nuclei. In the first case the total moment of the electron $\vec{j}$ and the operator $k=\beta (\vec{\sigma}\vec{l}+1)$ are conserved, and so the upper and the lower components of the electronic wave function will contain the real radial functions $f_\kappa$, $g_\kappa$ and the spherical spinors $X_{-|\kappa|, m_j} \equiv\Omega_{jlm_j}$, $X_{|\kappa|, m_j} \hm \equiv (\vec{\sigma} \vec{n} )\, \Omega_{jlm_j}$ (the definition of spherical harmonics and spinors follows Refs.~\cite{Bateman1953,Roenko2017c})
\begin{equation}\label{eq:7a}
\varphi = f_\kappa(r)\, X_{\kappa, m_j}\, , \qquad \chi = g_\kappa(r)\, X_{-\kappa, m_j}\, ,
\end{equation}
while for the energy levels with given $\kappa=\pm(j+\myfrac{1}{2})$ one obtains the spectral problem in the form of the following system of equations
\begin{equation}\label{eq:8a}
\begin{split}
\partial_r f_{\kappa} &+ \frac{1+\kappa}{r}\, f_{\kappa}  - \lambda\, \frac{Z\, \nu(r)}{r^2}\, f_{{\kappa}} = \left(1+\epsilon+ \alpha U(r)\right) g_{\kappa}\, ,\\
\partial_r g_{\kappa} &+ \frac{1-\kappa}{r}\, g_{\kappa}  + \lambda\, \frac{Z\, \nu(r)}{r^2}\, g_{{\kappa}} = \left(1-\epsilon- \alpha U(r)\right)f_{\kappa} \, ,
\end{split}
\end{equation}
where $\nu(r)=c(r)-r c'(r)$.

In the case of the quasi-molecule containing of the two nuclei, the centers of which are placed at the points $\vec{r}_{1,2} = \pm a \vec{e}_z$, only the projection $j_z$ is conserved, and the solutions of the Eqs.~\eqref{eq:6} with definite $m_j$ are represented now as the (truncated) expansions in spherical spinors
\begin{equation}\label{eq:7b}
\varphi = \sum_{\kappa = \pm 1}^{\pm K}  f_\kappa\, X_{\kappa, m_j}\, , \qquad \chi = \sum_{\kappa = \pm 1}^{\pm K}  g_\kappa\, X_{-\kappa, m_j}\, .
\end{equation}
As a result, the initial Eqs.~\eqref{eq:6} take the form of the system of $4K$ equations
\begin{equation}\label{eq:7}
\begin{split}
\partial_r f_{\kappa} &+ \frac{1+\kappa}{r}f_{\kappa}  + \lambda \sum_{\bar{\kappa}} M_{\kappa;\bar{\kappa}}(r)\, f_{\bar{\kappa}} = (1+\epsilon)g_{\kappa} + \alpha \sum_{\bar{\kappa}} N_{-\kappa;-\bar{\kappa}}(r)\, g_{\bar{\kappa}}\, ,\\
\partial_r g_{\kappa} &+ \frac{1-\kappa}{r}g_{\kappa}  - \lambda \sum_{\bar{\kappa}} M_{-\kappa;-\bar{\kappa}}(r)\, g_{\bar{\kappa}} = (1-\epsilon)f_{\kappa} - \alpha \sum_{\bar{\kappa}} N_{\kappa;\bar{\kappa}}(r)\, f_{\bar{\kappa}} \, ,
\end{split}
\end{equation}
where the coefficient functions $N_{\kappa;\bar{\kappa}}(r)$, $M_{\kappa;\bar{\kappa}}(r)$ are expressed via the multipole moments $U_n$, $V_n$ and 3j-symbols~\cite{Roenko2017c}
\begin{equation}\label{eq:15a}
\begin{split}
N_{\kappa;\bar{\kappa}}(r)  &= \sum_{n=0}^{2K} U_{n}(r)\, \langle X_{\kappa, m_j} | P_n(\cos \vartheta) | X_{\bar{\kappa}, m_j} \rangle, \\ M_{\kappa;\bar{\kappa}}(r)  &= \sum_{n=0}^{2K} \left( \partial_r + \frac{\kappa-\bar{\kappa}}{r} \right) V_n(r)\, \langle X_{\kappa, m_j} | P_n(\cos \vartheta) | X_{\bar{\kappa}, m_j} \rangle.
\end{split}
\end{equation}

All the Coulomb multipole moments $U_n$ are evaluated analytically in the model of nucleus charge used (see Ref.~\cite{Roenko2017c}). The systems of equations~\eqref{eq:8a} and~\eqref{eq:7} are 
solved by numerical methods; in the case of compact quasi-molecules ($d=2a\lesssim 100$~fm) the technique used converges quite fast with growing the cutoff $\kappa_{max}=K$ in the expansion~\eqref{eq:7a}, which allows one to determine the electronic levels energy with an accuracy of $10^{-6}\sim 10^{-7}$ electron mass~\cite{Roenko2017c}.


%
%
\section{Properties of the levels shifts due to $\Delta U_{AMM}$ }
\label{sec:2}
By solving the Eqs.~\eqref{eq:8a},~\eqref{eq:7} with and without the additional interaction $\Delta U_{AMM}$ 
one obtains the levels shifts caused by AMM.
 This shift in H-like atoms is a part of the self-energy contribution to the Lamb shift, which is usually represented in terms of the function $F_{nj}(Z\alpha)$, defined by~\cite{Mohr1998}
\begin{equation} \label{eq:21}
\Delta E^{SE}_{nj}(Z\alpha)=\frac{ Z^4 \alpha^5}{\pi n^3}\, F_{nj}^{SE}(Z\alpha) \, mc^2  \, .
\end{equation}

\begin{figure*}
\begin{minipage}[b]{.485\textwidth}\label{pic:4a}
\includegraphics[width=\textwidth]{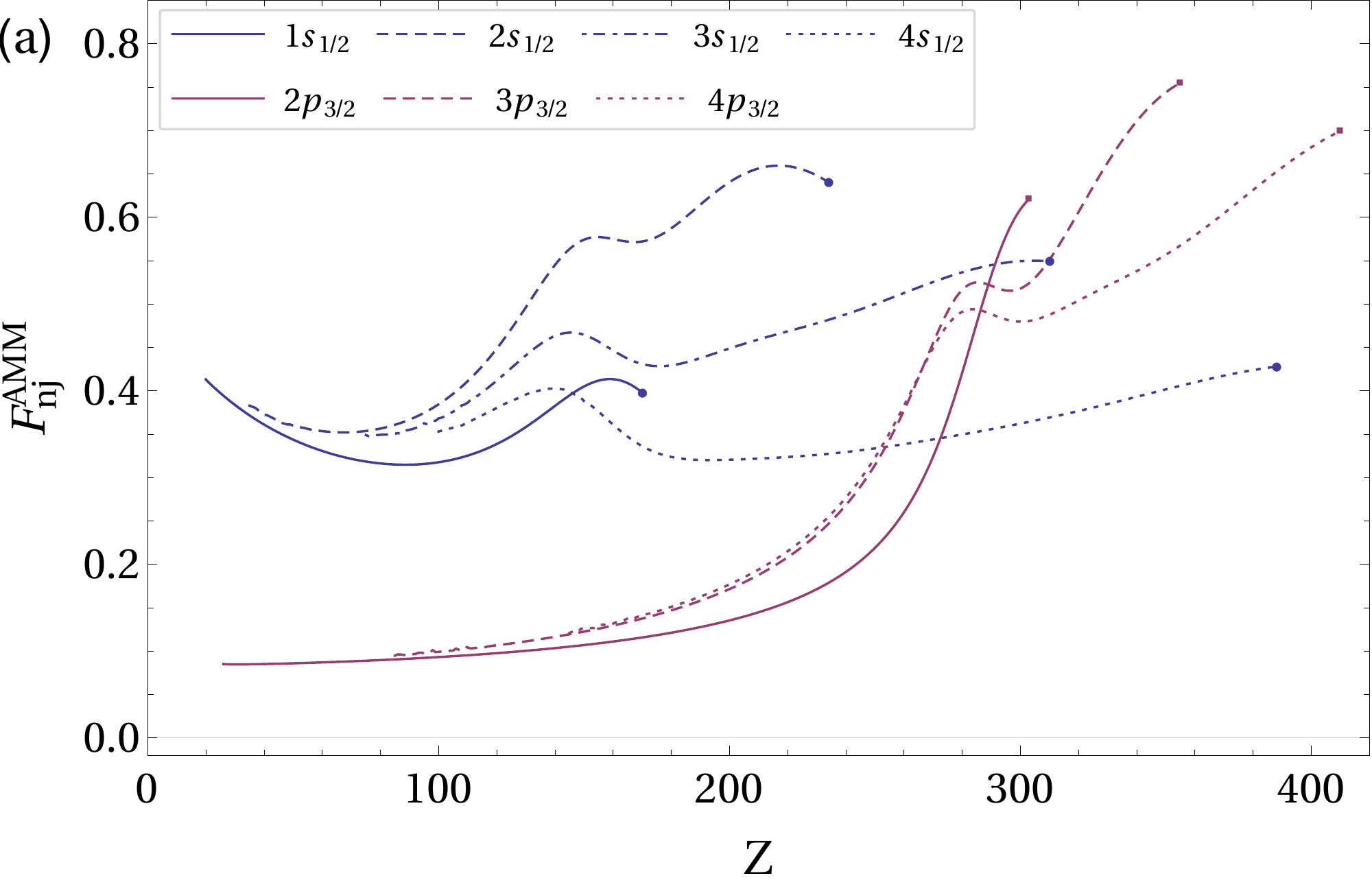} 	%
\end{minipage}
\hfill
\begin{minipage}[b]{.495\textwidth}\label{pic:4b}
\includegraphics[width=\textwidth]{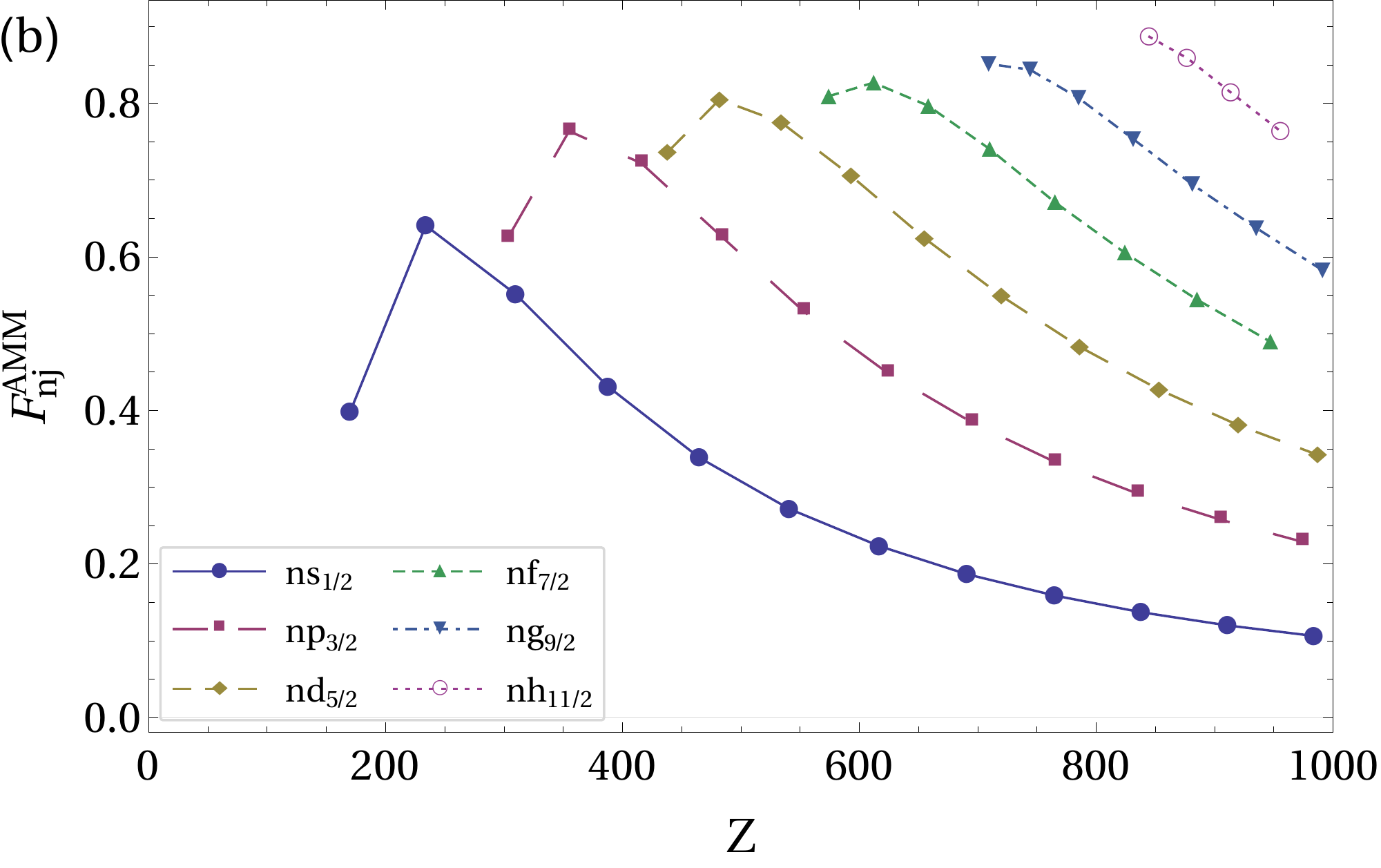} 	%
\end{minipage}
\begin{minipage}[b]{.485\textwidth}
\includegraphics[width=\textwidth]{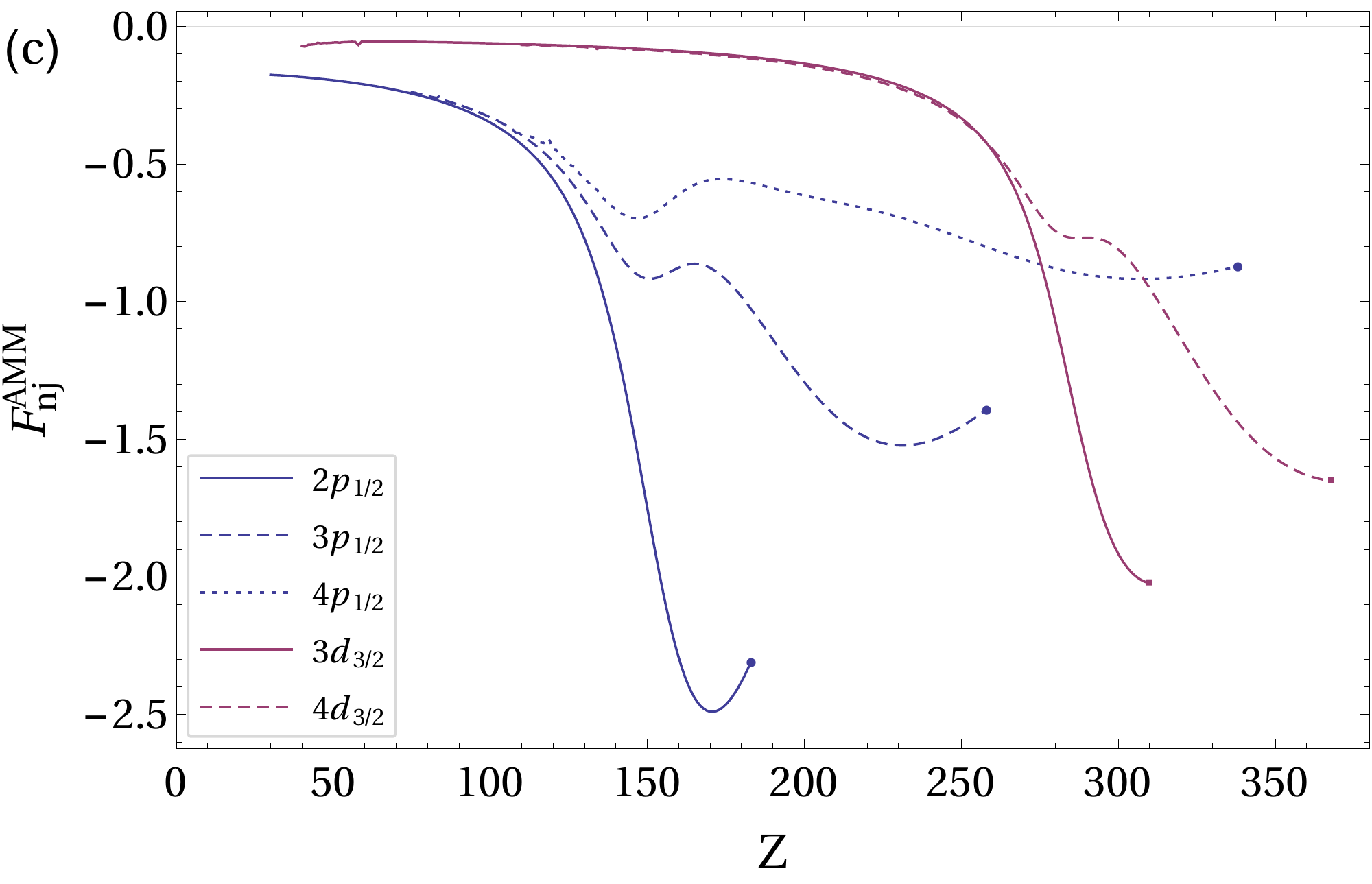} 	%
\end{minipage}
\hfill
\begin{minipage}[b]{.495\textwidth}
\includegraphics[width=\textwidth]{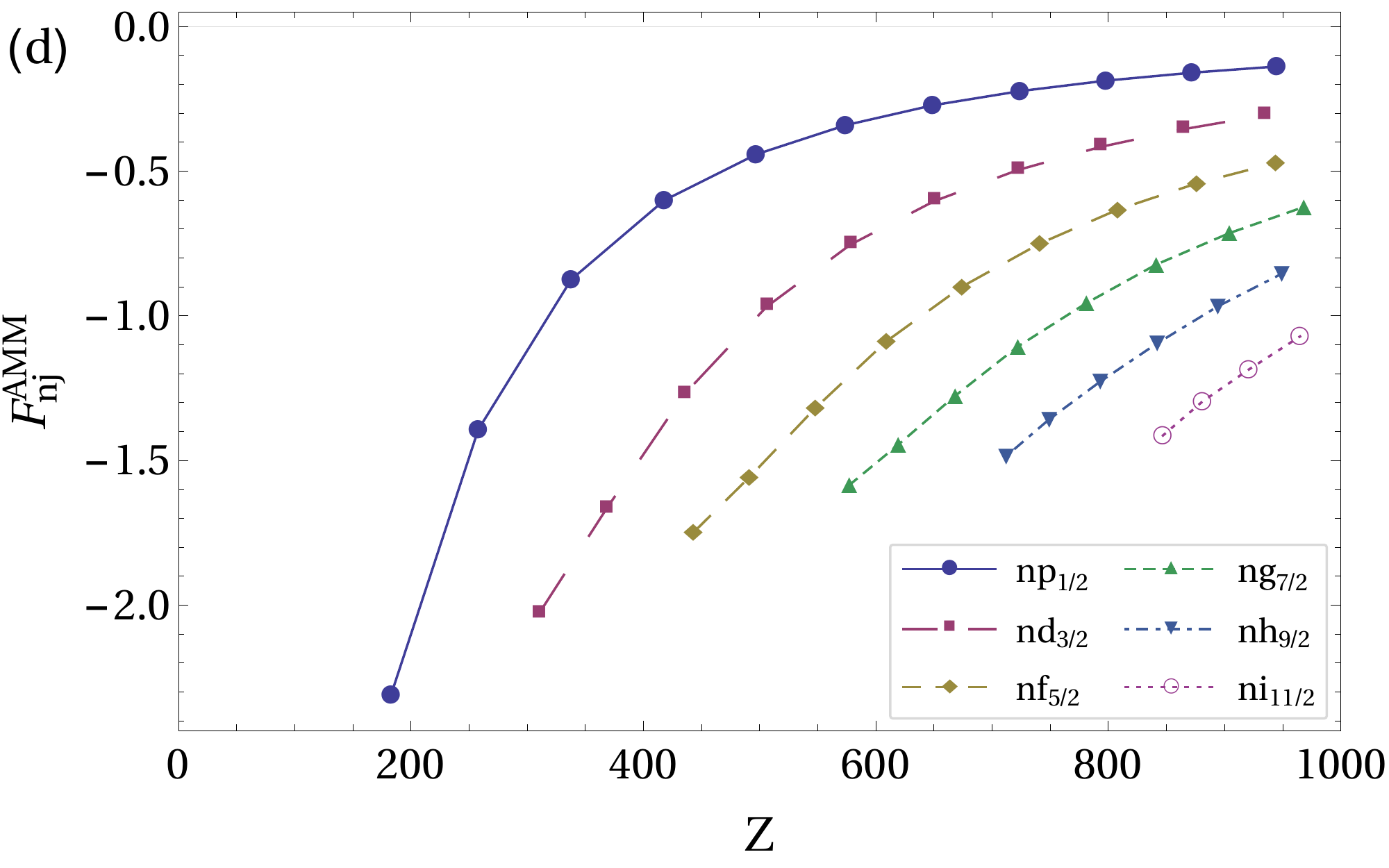} 	%
\end{minipage}
\caption{The shift of the levels in H-like atom due to $\Delta U_{AMM}$ in terms of $F_{nj}^{AMM}(Z\alpha)$ as a function of the nuclear charge~$Z$ for levels with $n\leq 4$ and $j=\myfrac{1}{2}, \myfrac{3}{2}$ (a),(c); $F_{nj}^{AMM}(Z_{cr,nj} \alpha)$ for the shifts of the levels located near the threshold of the lower continuum in the region $Z < 1000$ (b),(d). The separate groups of points in Figs.~(b),(d) correspond to the series of the levels with fixed $lj$ and various $n$.\vspace{-0em}
}
\label{fig:1}\label{fig:2} 
\end{figure*}

\begin{figure*}
\begin{minipage}[b]{.49\textwidth}\label{pic:2N-1s-F}
\includegraphics[width=\textwidth]{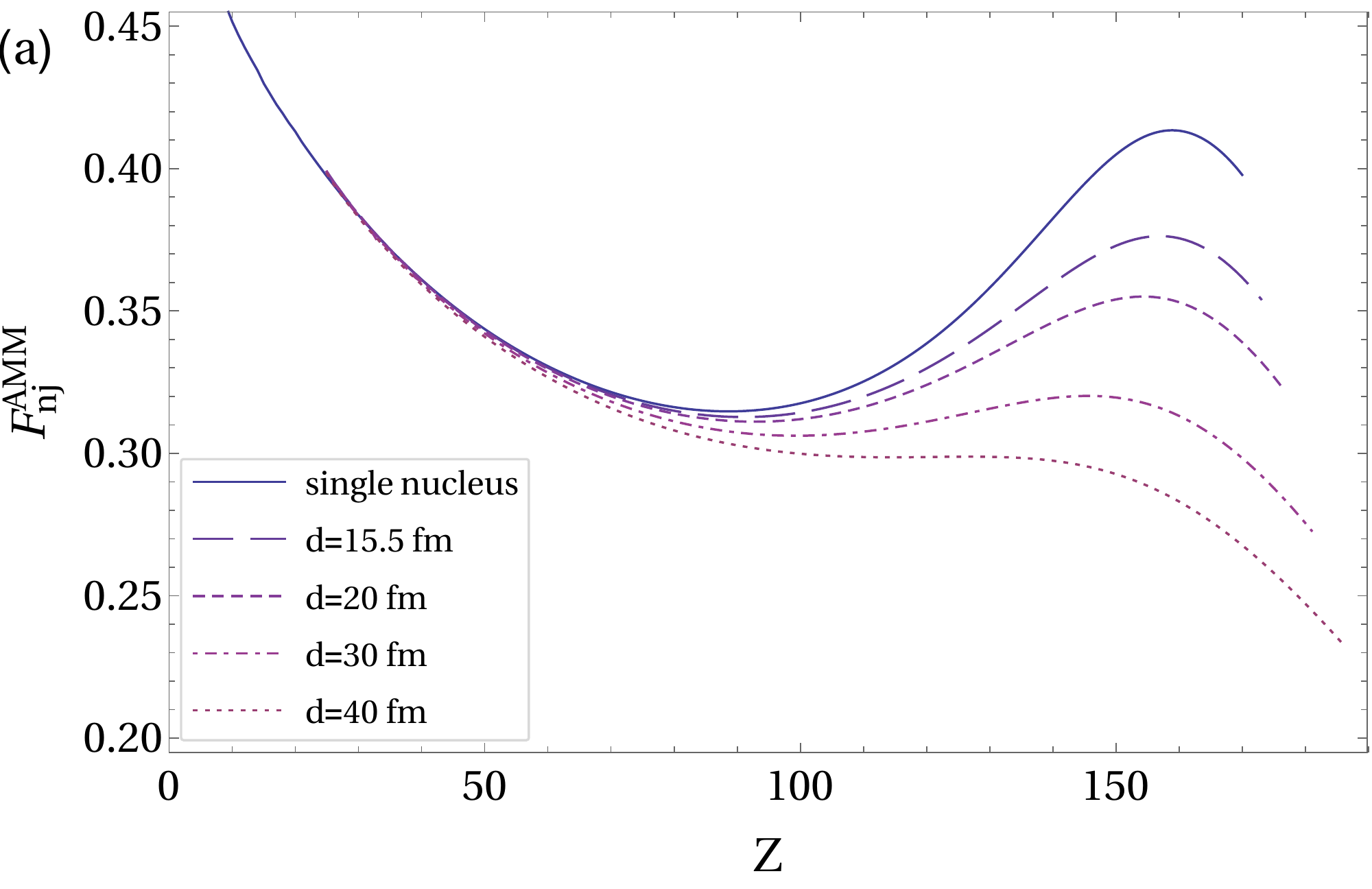}
\end{minipage}
\hfill
\begin{minipage}[b]{.49\textwidth}\label{pic:2N-2p-F}
\includegraphics[width=\textwidth]{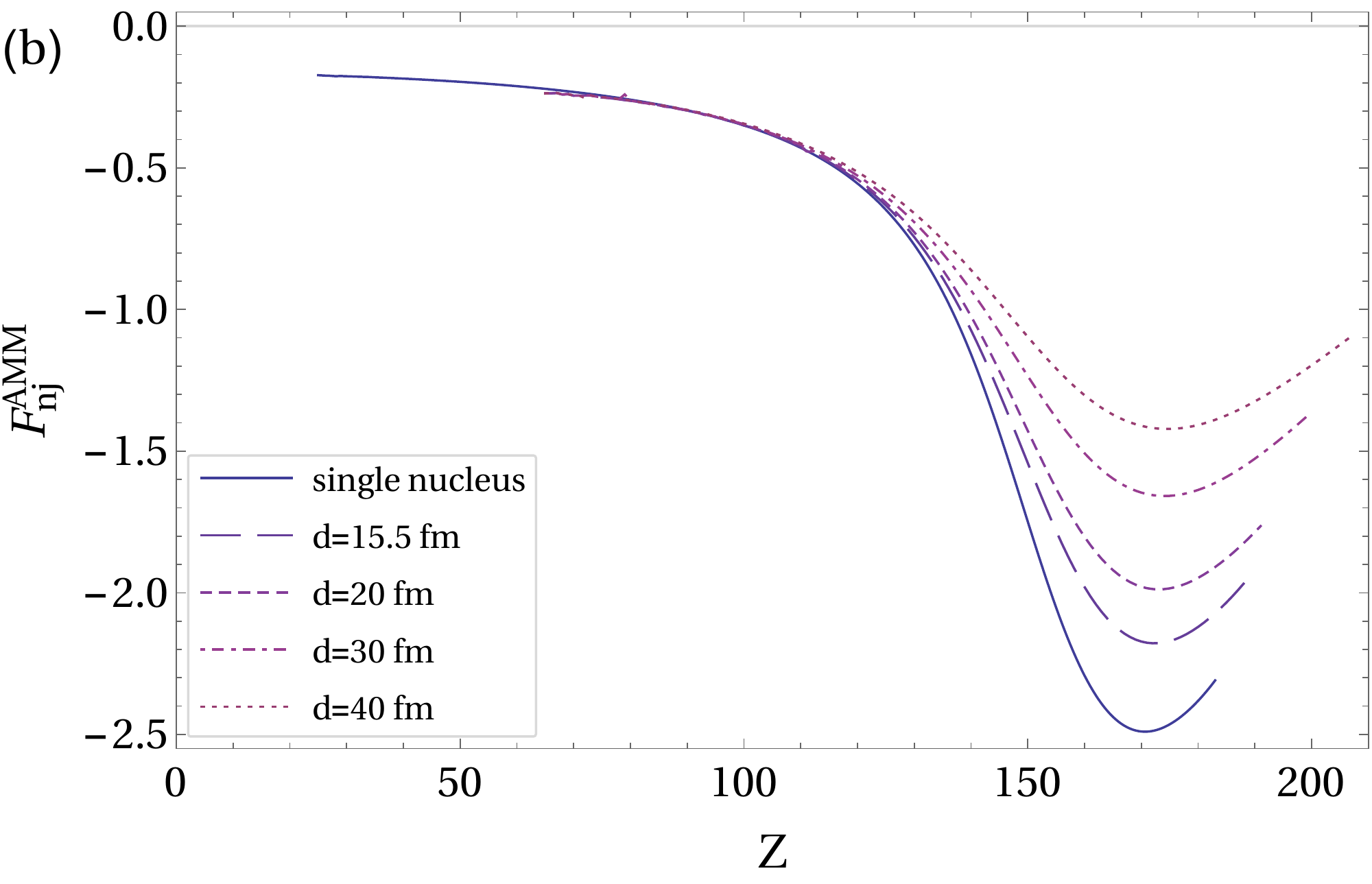}
\end{minipage}
\caption{The function $F_{nj}^{AMM}$ for the contribution from $\D U_{AMM}$ as a function of total charge  $Z$ for the lowest even (a) and odd (b) electronic levels in an H-like atom (solid) and in a symmetrical two-nuclei quasi-molecule for the fixed internuclear distances (other).\vspace{-0em}}\label{fig:3} 
\end{figure*}

In the Figs.~\ref{fig:1}(a) and~\ref{fig:1}(c) the function $F_{nj}^{AMM}(Z\alpha)$ is plotted for a number of the lower electronic levels in the H-like atom. The right end point of each curve in Figs.~\ref{fig:1}(a) and~\ref{fig:1}(c) corresponds to the shift of the level $nlj$ in H-like atom with critical nuclear charge $Z_{cr,nj}$ (in this system the level is located at the threshold $\epsilon=-1$). There follows from Fig.~\ref{fig:1} that the levels shifts due to electronic AMM have a non-monotonic dependence on $Z$ with local peaks, when the levels approach the threshold of the lower continuum. Nevertheless, the magnitude of the levels shifts near $\epsilon = -1$ decreases with growing $Z$ for each series of the levels (see Figs.~\ref{fig:2}(b) and~\ref{fig:2}(d)). In the quasi-molecule the shift due to $\Delta U_{AMM}$ decreases quite rapidly with increasing internuclear distance, moreover, the shift near the threshold also falls down with enlarging the distance between nuclei (both in absolute units and in units of $Z^4\a^5/\pi n^3$), although the total charge of the nuclei $Z_{cr}$ becomes greater (see Fig.~\ref{fig:3}).


%
%
\section{Conclusion}
\label{sec:con}
Thus, we have shown how the electronic levels behave due to $\Delta U_{AMM}$ in the one-center and two-center systems with subcritical and supercritical charge. It should be noted that these results are completely non-perturbative in $Z\alpha$ and (partially) in $\alpha/\pi$, since the latter enters the coupling constant of $\Delta U_{AMM}$.

And although the shift due to AMM  is just a part of the whole radiative correction to the binding energy\footnote{For example, the $1s_{1/2}$ level in H-like atom with critical charge $Z=170$ is shifted by $\Delta E^{AMM}_{1s}\simeq 1.12$~keV due to the interaction~\eqref{eq:f3}, which is approximately the tenth part of the whole self-energy shift $\Delta E^{SE}_{1s}\simeq 11.0$~keV~\cite{Mohr1982}.}, the key-point here is that the behavior of $F_{nj}^{AMM}(Z\alpha)$ qualitatively reproduces the behavior of $F_{nj}^{SE}(Z\alpha)$ in subcritical region for a number of the lower electronic levels~\cite{Roenko2017a}. So there appears a natural assumption, that in the overcritical region the decrease with the growing  $Z$  and the size of the system of Coulomb sources should take place also for the total self-energy contribution to the levels shift near the threshold of the lower continuum, and hence, for the other radiative QED effects with virtual photon exchange.
Thus, one may expect, that the non-linear decline of the vacuum energy for $Z>>Z_{cr}$, where the contribution from the fermionic loop plays the main role~\cite{Davydov2017, Sveshnikov2017, Davydov2018a}, can not be compensated by the contribution from the radiative corrections, but this statement requires  more thorough further verification.

\bibliography{biblio/AMMp}
%

\end{document}